\documentstyle[12pt]{article}
\begin{document}
\def\today{\space\number\day\ \ifcase\month\or
January\or February\or March\or April\or May\or June\or July\or
August\or September\or October\or November\or December\fi
\ \number\year}
\overfullrule=0pt  
\def\mynote#1{{[{\it NOTE: #1}]}}
\def\fEQN#1#2{$$\hbox{\it #1\hfil}\EQN{#2}$$}
\def\Acknowledgements{{\bigskip\leftline
{{\bf Acknowledgments}}\medskip}}

\begin{titlepage}

\begin{flushright}
DFTUZ/96-10\\  

\end{flushright}

 \vspace{0cm}
 
\begin{center}
{\large\bf  Effective Field Theory of pure Gravity\\    
and the Renormalization Group  }

 \vspace{0.4cm}
 
{\bf Mario Atance$^{\ast}$}\footnote{E-mail atance@posta.unizar.es}  and
{\bf Jos\'e~Luis~Cort\'es$^\dagger$}\footnote{E-mail cortes@posta.unizar.es}  

 \vspace{0.2cm}

{\sl Departamento de F\'{\i}sica Te\'orica,\\ 
Universidad de Zaragoza,
50009 Zaragoza, Spain.}

 \vspace{0.2cm}

\centerline{ March 1996}

\end{center}
\vspace{0.2cm}
\begin{abstract}

The general structure of the renormalization group equations for
the low energy effective field theory formulation of pure gravity
is presented. The solution of these equations takes a particular 
simple form if the mass scale of the effective theory is much smaller 
than the Planck mass (a possibility compatible with the renormalization 
of the effective theory). A theory with just one free renormalized 
parameter is obtained when contributions suppressed by inverse powers 
of the Planck mass are neglected.
\end{abstract}

\end{titlepage}
\hfill

The general principles of quantum mechanics and special relativity
lead to identify any relativistic quantum theory at low energy as 
a quantum field theory~\cite{Weinberg I}. This point of view of
field theories as effective low energy theories has important 
consequences on the understanding of general problems in quantum
field theory such as renormalization~\cite{Polchinski}. The
finite number of interactions of a renormalizable field theory
are just the dominant terms at low energy of an effective field
theory with an infinite number of interactions (usually all the 
terms compatibles with some symmetry principles). 

On the other hand the present incompatibility of general relativity 
and quantum mechanics~\cite{Isham} can be translated into the 
need for restricting the field theoretical formulation of gravitation 
to energies below the Planck mass. In field theory this 
incompatibility appears through the non-renormalizability of the
theory of a spin 2 massless interacting particle. But in the framework 
of effective field theories renormalizability is not a consistency
requirement and reliable perturbative predictions after 
renormalization can be made~\cite{EFT}. Then one can consider the
effective field theory of an interacting spin 2 massless particle
as the low energy field theoretic formulation of pure gravity. 
This point of view has been recently advocated by 
J.F.Donoghue~\cite{Donoghue} who shows how some large distance 
quantum gravitational effects can be derived within this framework.
At high enough energies new interactions and new degrees of freedom 
will be required in a new theory based on some unknown general
principles~\cite{Isham}.

Perturbative renormalization in effective field theories can be 
done along the same lines of the standard renormalization of 
renormalizable theories. The example of the effective
field theory that reproduces the low energy effects
of a renormalizable field theory with a very 
heavy field can be used to illustrate the main ideas~\cite{Manohar}.
A well defined effective field theory expansion in inverse
powers of the heavy mass is obtained if an appropriate 
renormalization scheme is used in the effective theory (mass 
independent substraction scheme).

The one loop~\cite{t'Hooft-Veltman} and two 
loop~\cite{Goroff-Sagnotti} calculations of ultraviolet divergences 
in Einstein theory of gravity using dimensional regularization can 
be understood as the perturbative determination  of the
counterterms invariant under general covariant transformations
with four and six derivatives which are required 
in the renormalization of the effective low energy theory of pure 
gravity. The loop expansion corresponds to the expansion in 
derivatives and the renormalized theory has an
infinite number of terms. This expansion can be seen as the effective
field theory expansion in inverse powers of the mass scale of the 
theory.

The purpose of this paper is to argue that, despite of the fact that one 
has in general an infinite number of free parameters,
something can be learned about quantum gravity from its
effective field theory formulation. The idea is to look at    
the information contained in the general structure of the
renomalization group equations of the theory.

The general expression for the action of the effective field 
theory of gravitation can be writen, using the invariance under
general covariant transformations, in the following form

\begin{equation} S=\int d^{4}x \, {\sqrt {-g}} \,
{M^{2} \over \alpha_{1}^{2}} \lbrack R + 
{{\vec \alpha}_{2} \over M^{2}} \, {\vec R}^{(2)} \, + \,
{{\vec \alpha}_{4} \over M^{4}} \, {\vec R}^{(4)} \, + \, ...
\rbrack  \label{action} .\end{equation}

\noindent A constant term inside the brackets has not been included by
using the experimental information that the cosmological constant
must be very small (later we will see how the structure of the
renormalization group would be affected in the presence of such a
term). The coefficients $ \alpha_{1}$, $\vec \alpha_{2}$, 
$\vec \alpha_{4}$,... are dimensionless parameters, $ R $ is the
scalar curvature, $\vec {R^2} $ is a vector with the three different
invariants built out of two Riemann tensors as components, 
the different invariants with three Riemann tensors or two Riemann 
tensors and two derivatives are the components of the vector 
$\vec {R^4}$ and so on\footnote{One can see that the action 
(\ref{action}) has redundant terms that could be eliminated by using 
a non-linear redefinition of fields. We will not make use of this 
possibility to eliminate some terms in the action in this paper.}.
The dimensionality of the different terms in the action 
fixes the power dependence on the mass scale $M$ of the 
effective theory. The Newtonian limit of this action gives 
${\alpha_1}^2 = 16 \pi ( {M \over {M_{Pl}}})^2 $ where $M_{Pl}$ is
the Planck mass. 

A perturbative analysis of the action (\ref{action}), based on the 
decomposition of the metric
\begin{equation} g_{\mu \nu} = \eta_{\mu \nu} + {\alpha_1 \over
M} h_{\mu \nu} \,\,\,\,\,\,  \eta_{\mu \nu} = diag(1, -1, -1, -1), 
\end{equation}
can be done~\cite{t'Hooft-Veltman,Goroff-Sagnotti,Veltman} 
by using the standard methods of gauge theories. A very 
important tool in handling infinities in this analysis is 
dimensional regularization~\cite{DR} and the minimal 
substraction scheme~\cite{t'Hooft}. The
standard derivation, in perturbatively renormalizable theories, of 
the renormalization group in this renormalization scheme~\cite{Gross}
can be translated to an effective field theory. When
this renormalization scheme is applied to the action (\ref{action}) 
one finds a set of bare parameters 
$ {\vec \alpha}^{B} = (\alpha_{1}^{B}, {\vec \alpha_{2}}^{B},...)$
in one to one correspondence with the dimensionless parameters
in the action (\ref{action}). The expressions for the bare 
parameters in terms of the renormalized parameters will have poles 
when $ \epsilon \rightarrow 0 $ (dimension $D = 4 - \epsilon $).
From the independence of the bare parameters on the renormalization
scale $\mu$, one concludes that any change of $\mu$ must be
equivalent to a change in the renormalized parameters. The 
renormalization group equations express this fact,

\begin{equation} \mu {d \alpha_{1} \over d \mu} \,=\, 
\beta_{1} ({\vec \alpha}) \,\,\, , \,\,\,\,
\mu {d {\vec \alpha}_{2n} \over d \mu} \,=\,
{\vec \beta}_{2n} ({\vec \alpha}) 
\label{RGE} \,\,\,.\end{equation}

\noindent The renormalization group $\beta$ functions 
${\vec \beta}_{2n}$ are vectors with as many components as 
${\vec \alpha}_{2n}$, i.e., the number of covariant terms with 
dimensionality $2n + 2$. Each component of ${\vec \beta}_{2n}$
is a polinomial in the dimensionless parameters compatible with 
the homogeneity condition,

\begin{equation} \beta_{2n}^{(i)} ( \lambda^{k} {\vec \alpha}_{k} )
\, = \, \lambda^{2n} \beta_{2n}^{(i)} ({\vec \alpha})
\label{rghc} \,\,\,\,.\end{equation}

\noindent This is a consequence of dimensional arguments together 
with the presence of a single mass scale $M$ (the dependence on
the renormalization scale $\mu$ is logarithmic). From (\ref{rghc})
one can see that the $\mu$ dependence of ${\vec \alpha}_{2n}$ is
fixed by a finite number of parameters ${\vec \alpha}_{k}$, $k < 2n$.

This ``triangular" structure of the renormalization group equations
dissappears if a cosmological constant term is added in (\ref{action}).
In the parametrization we are using the cosmological term would 
correspond to a term $\alpha_{-2} \, M^2$ inside the brackets in 
(\ref{action}). The only modification in the renormalization group
equations (\ref{RGE}) is the addition of one equation for the new
dimensionless parameter $\alpha_{-2}$ and the presence of new
contributions in the ${\vec \beta}$ functions depending on 
$\alpha_{-2}$. The homogeneity conditions (\ref{rghc}) are still
valid but, due to the presence of $\alpha_{-2}$, all the dimensionless
parameters in the action (\ref{action}) appear in every renormalization
group equation. One consequence of these generalized homogeneity
conditions is that $\beta_{-2}$ is proportional to $\alpha_{-2}$
and then a vanishing cosmological constant is consistent with
the renormalization group equations~\cite{Atance-Cortes}. From now on 
we will omit the cosmological term and we will work with the effective
action (\ref{action}).

The next step is to consider the possibility of eliminating more
parameters in the renormalized theory or in other words the 
identification of relations among the renormalized parameters 
${\vec \alpha}_{k}$
compatible with their renormalization scale dependence. This can be
seen as an example of the program of coupling constant reduction
initiated in~\cite{Oehme-Zimmerman} and discussed in more general
terms in ~\cite{Perry-Wilson}. In the present case, the dimensional
arguments which lead to the simple ``triangular" structure of the
renormalization group equations put strong limitations on the 
possible relations among parameters. A reduced parameter 
$\alpha^{(i)}_{2n}$ will be given by
   
\begin{equation} 
\alpha^{(i)}_{2n} = {r}^{(i)}_{2n} ( {\vec \alpha} )
\label{reduction}\,\,\,\,\, ,\end{equation}
where ${r}^{(i)}_{2n}$ is a polinomial in ${\vec \alpha}$ with
the homogeneity condition

\begin{equation} r_{2n}^{(i)} ( \lambda^{k} {\vec \alpha}_{k} )
\, = \, \lambda^{2n} r_{2n}^{(i)} ({\vec \alpha})
\label{rhc} \,\,\,.\end{equation}

\noindent Once more one has a very simple structure in the reduction 
with any reduced parameter $\alpha^{(i)}_{2n}$ depending on 
${\vec \alpha}_{k}$, $k < 2n$ exclusively. From the renormalization
group equations for $\alpha_{1}$  and ${\vec \alpha}_{2}$

\begin{equation} \mu {d \alpha_{1} \over d \mu} \,=\, 0
\label{RGE1} \end{equation}

\begin{equation} \mu {d \alpha_{2}^{(i)} \over d \mu} \,=\,
{b_{2}^{(i)} \over (4 \pi)^{2}} \alpha_{1}^{2}
\label{RGE2} \end{equation}
one has a consistent reduction 

\begin{equation} \alpha_{2}^{(i)} \, = \, b_{2}^{(i)} \alpha_{2} 
+ c_{2}^{(i)} \alpha_{1}^{2}
\label{re2},\end{equation}
where the coefficients $c_{2}^{(i)}$ in the second term are 
completely free as a consequence of the trivial renormalization 
of $\alpha_{1}$. The independent dimensionless
parameter $\alpha_{2}$, corresponding to terms quadratic in the
curvature, has a $\mu$ dependence given by 

\begin{equation} \mu {d \alpha_{2} \over d \mu} \,=\,
{1 \over (4 \pi)^{2}} \, \alpha_{1}^{2} 
\label{RRGE2} \,\,\,.\end{equation}

\noindent If the same arguments are applied to the renormalization
group equations for ${\vec \alpha}_{4}$ one finds a reduction

\begin{equation} \alpha_{4}^{(i)} \, = \, b_{4}^{(i)} \alpha_{2}^{2}
+ b_{4}^{'(i)} \alpha_{2} \alpha_{1}^{2} 
+ c_{4}^{(i)} \alpha_{1}^{4} 
\label{re4} \,\,\,\,.\end{equation}

\noindent The coefficients of the first two terms, $b_{4}^{(i)}$ and
$b_{4}^{'(i)}$ are fixed by the consistency of the renormalization 
group equations, but once more there is no restriction on the 
coefficient $c_{4}^{(i)}$ of the last ($\mu$-independent) term.
The generalization of these results to ${\vec \alpha}_{2n}$ is
obvious. 

The presence of an arbitrary constant in any reduction
equation, which is a consequence of 
the $\mu$-independence of $\alpha_{1}$, makes the reduction of
parameters trivial in this case. It just corresponds to a 
parametrization of the most general solution of the renormalization
group equations in terms of the arbitrary constants ${\vec c}_{2n}$
and one has an infinite number of renormalized parameters that can
be fixed independently at a given scale. 

If all the dimensionless parameters are $\alpha^{(i)}_{k} \sim 1$ 
at the scale $M$ then

$$M^{2} = {\alpha_{1}^{2} \over {16 \pi}} \, 
M_{Pl}^{2} \sim M_{Pl}^{2}$$
and one has an effective theory of gravity with the Planck mass
playing the role of the scale of the theory. For energies 
$E \ll M_{Pl}$ (long distance physics), one has a dominant 
contribution independent of ${\vec \alpha}_{2n}$. At energies
$E \sim M_{Pl}$ the contributions depending on the parameters
${\vec \alpha}_{2n}$ become comparable to the dominant term 
at lower energies. Predictibility is lost because one goes 
beyond the domain of applicability of the effective field
theory of gravity.

An alternative to the most natural case discussed previously
corresponds to keep the dimensionless parameters
$\alpha^{(i)}_{2n} \sim 1$ but to assume that $\alpha_{1} \ll 1$.
If one takes the dominant term in the reduction equations then 
one has a ``real" reduction of couplings and the theory involves 
in this approximation just one free dimensionless parameter
$\alpha_{2}$ together with the mass scale $M$ of the theory (or
equivalently $\alpha_{1}$ which is the mass scale in units of 
the Planck mass). The explicit form of the first few terms in the 
action can be obtained from previous calculations of one and two
loop divergences in quantum gravity.

The one loop result~\cite{t'Hooft-Veltman,Goroff-Sagnotti} for the 
counterterms

\begin{equation} {\cal L}_{c.t.}^{(1)} \, = \,
{{\sqrt {-g}} \over (4 \pi)^{2} \epsilon} \lbrack {1 \over {120}} 
R^{2} + {7 \over {20}} R_{\mu \nu} R^{\mu \nu} - 
{53 \over {180}} \epsilon^{\alpha \beta \gamma \delta}
\epsilon_{\mu \nu \rho \sigma} {R^{\mu \nu}}_{\alpha \beta}
{R^{\rho \sigma}}_{\gamma \delta} \rbrack
\label{counterm} \end{equation}

\noindent gives, after reduction of couplings

\begin{equation} {\vec \alpha}_{2} \, {\vec R}^{(2)} \, = \,
 \alpha_{2} \lbrack {1 \over {120}} R^{2} +
{7 \over {20}} R_{\mu \nu} R^{\mu \nu} - 
{53 \over {180}} \epsilon^{\alpha \beta \gamma \delta}
\epsilon_{\mu \nu \rho \sigma} {R^{\mu \nu}}_{\alpha \beta}
{R^{\rho \sigma}}_{\gamma \delta} \rbrack
\label{redefaction} .\end{equation}

\noindent The next term in the reduced action will be obtained by 
using the approximation 
$\alpha_{4}^{(i)} \approx b_{4}^{(i)} \, \alpha_{2}^{2}$,
for the coefficients of the ten 
independent terms of dimensionality six, 
where the coefficients $b_{4}^{(i)}$ can be read from the
${1 \over \epsilon}$ two loop counterterms~\cite{Goroff-Sagnotti}.
The determination of ${\vec \alpha}_{6}$ would require a three
loop calculation and so on. 

A generic term in the energy expansion associated to the 
effective theory $\alpha_{2n}^{(i)} \left({E \over M}\right)^{2n}$
becomes, after reduction of couplings, 
$\left({\alpha_{2} \over \alpha_{1}^{2}}\right)^{n} \,\,
\left({E \over M_{Pl}}\right)^{2n}$ and then the energy expansion
involves a single free dimensionless parameter
${\alpha_{2} \over \alpha_{1}^{2}}$. If one considers the energy
expansion of the effective theory in the general case without the 
approximation $\alpha_{1} \ll 1$, the reduction of couplings, which
is simply a particular parametrization of the action, corresponds
to the identification of a double expansion in powers of
${E \over M}$ and ${E \over M_{Pl}}$. The approximate ``real"
reduction discussed in this paper is a consequence of a 
property of this double expansion. The infinite number of 
parameters of the effective action do not appear in the coefficients
of terms $\sim \left({E \over M_{Pl}}\right)^{0} \,
\left({E \over M}\right)^{2n}$ in the double expansion. These 
coefficients depend only on two parameters
$\alpha_{1}$ and  $\alpha_{2}$.

If the discussion of this paper is extended to an effective 
theory of gravity, including matter fields and non-gravitational
interactions, and the structure of the energy expansion is not
modified, there can be interesting
implications of a mass scale of the effective theory $M$  much 
smaller than the Planck mass.

\bigskip

\bigskip\leftline{{\bf Acknowledgments}}\medskip
This work was partially supported by the CICYT (proyecto AEN 94--0218).  
The work of M.A. has been supported by a DGA fellowship.
 
\vfill
\eject

\newpage

\end{document}